\documentclass[12pt]{article}
\usepackage[numbers]{natbib}
\usepackage{a4}
\usepackage{amsmath}
\usepackage{amssymb}
\usepackage{latexsym}
\usepackage{amssymb}
\def \dd{{\mathrm{d}}}

\def \pd{\partial}

\def \tl#1{\overset{\kern 1pt\circ}{#1}}
\def \TL#1{\overset{\kern -3pt \circ}{#1}}
\def \TLL#1{\overset{\kern -7pt \circ}{#1}}

\def \BPhi{\boldsymbol{\Phi}}

\def\quer{\!\!\!\!\!\!\!\nearrow}

\begin{document}
\title{{\bf On microcontinuum field theories:\\
the Eshelby stress tensor and incompatibility conditions}}
\author{Markus Lazar~$^\text{a,}$\footnote{
Corresponding author. {\it E-mail address:} lazar@fkp.tu-darmstadt.de (M.~Lazar).}\; ,
G{\'e}rard A.~Maugin~$^\text{b,}$
\\ \\
${}^\text{a}$
Emmy Noether Research Group,\\
        Department of Physics,\\
        Darmstadt University of Technology,\\
        Hochschulstr. 6,\\      
        D-64289 Darmstadt, Germany\\
${}^\text{b}$ 
Laboratoire de Mod{\'e}lisation en M{\'e}canique, UMR 7607 CNRS,\\
        Universit{\'e} Pierre et Marie Curie,\\
        4 Place Jussieu, Case 162,\\    
        F-75252 Paris Cedex 05, France
}

\date{\today}    
\maketitle

\begin{abstract}
We investigate linear theories of incompatible micromorphic elasticity,
incompatible microstretch elasticity, incompatible micropolar elasticity
and the
incompatible dilatation theory of elasticity (elasticity with voids). 
The incompatibility conditions and Bianchi identities are derived and discussed.
The Eshelby stress tensor (static energy momentum) is calculated for 
such inhomogeneous media with microstructure.
Its divergence gives the driving forces 
for dislocations, disclinations, point defects and inhomogeneities
which are called configurational forces. 
\\

\end{abstract}
\vspace*{6mm}

\section{Introduction}
\setcounter{equation}{0}
In this paper we consider microcontinuum field theories like 
micromorphic elasticity, microstretch elasticity, micropolar elasticity and
dilatation elasticity.
The aim of this paper is to derive the incompatibility conditions, the 
Eshelby stress tensor~\citep{Eshelby51,Eshelby75} and the configurational forces, which are still
missing for such materials, except the micropolar elasticity, 
in the literature.
The micromorphic continuum theory was essentially developed by 
\citet{ES64,Eringen90,Eringen92,Eringen99,Mindlin64,Toupin64,Maugin70}
and \citet{Germain}.
A micromorphic theory can be used to model materials with 
microstructure.
A micromorphic continuum, as we consider, is built up from
particles which possess an inherent orientation.
Such a micromorphic theory is quite interesting to study due
to technical and mathematical reasons.

A micromorphic elastic solid possesses twelve degrees of freedom:
three translational ones and nine ones of the microstructure (rotation,
shear, dilatation). 
The micromorphic theory can be reduced to
microstretch elasticity, micropolar elasticity and 
dilatation elasticity. 
A microstretch elastic solid has seven degrees of freedom: 
three for translation, three for rotations and one for stretch.
Micropolar media possess three rotations and three translations.
In dilatation elasticity the microstructure has only one degree of 
freedom of dilatation in addition to the three classical translation
degrees of freedom. 
In presence of defects such theories are incompatible microcontinuum
field theory. The reason for the defects are plastic parts of the 
total strain fields of the theory under consideration.

Only for micropolar elasticity the mathematical 
expressions for
the Eshelby stress tensor and the configurational forces are already known.
They are given and discussed by~\citet{Kluge,Kluge2,Jaric,PS,NZ,Maugin98,LM}
and \citet{LK06}.
The same is true for the incompatibility equations defining the defect
fields (dislocations and disclinations).  
Quite recently, \citet{LA06} have given the Eshelby stress tensors for
compatible microstretch and micromorphic elasticity.

The aim of this paper is to close this lack in the literature.
We start our study with the theory of micromorphic elasticity. 
We derive the incompatibility equations which are defining the defect 
fields (dislocations, disclinations and (intrinsic) point defects), the 
Bianchi identities as conservation laws, the Eshelby stress tensor 
by means of (rigid) translational invariance and the 
configurational forces as divergence of the Eshelby stress tensor. 
Then we reduce such equations for microstretch elasticity, micropolar elasticity
and dilatation elasticity from the micromorphic ones.

\section{Theory of micromorphic elasticity}
\subsection{Field equations}
\setcounter{equation}{0}
We consider the linear theory of micromorphic elasticity 
(see, e.g.,~\citet{Eringen92,Eringen99}).
Accordingly, we have to assume that a micromorphic continuum, which
represents a continuum with microstructure, 
has additional deformational degrees of freedom.
The basic idea is to attach to each point of a three-dimensional
continuum three (or in general $n$) 
linearly independent vectorial `directors'.
In addition to the classical deformation, sometimes called macro-deformation,
the state of the continuum is specified by the deformation of the directors.
For a crystalline solid the directors could be interpreted 
as the position of the atoms with respect to the center of mass of the primitive cell
if the primitive cell contains more than one atom. In liquid crystals the 
director fields characterize the molecules.

In micromorphic elasticity 
the equilibrium equations for the physical fields (force stress $\sigma_{ij}$,
micro-stress $s_{ij}$,  
micro-hyperstress tensor $\mu_{ijk}$) are given by:
\begin{align}
\label{EC1}
&\pd_j\sigma_{ij}+f_i=0 ,\\
\label{EC2}
&\pd_k\mu_{ijk}+\sigma_{ij}-s_{ij}+l_{ij}=0,
\end{align}
where $f_i$, $l_{ij}$, are the body force and the body moment tensor, 
respectively.
The hyperstress tensor can be decomposed into
its antisymmetric part\footnote{We use the notation of~\citet{schouten2}. 
Symmetrization over two indices is denoted by
parentheses, $A_{(ij)}:=(A_{ij}+A_{ji})/2!$, antisymmetrization
by brackets, $B_{[ij]}:=(B_{ij}-B_{ji})/2!$.}, 
its trace, and its symmetric traceless
part with respect to its first two indices $i$ and $j$
\begin{align}
\mu_{ijk}=
\underbrace{\mu_{[ij]k}}_{\text{moment stress}}
+\ 
\frac{1}{3}\,\delta_{ij}
\underbrace{ \mu_{k}}_{\text{dilatation stress}}
+\underbrace{
{{\mu\quer}}{}_{(ij)k}}_{\text{shear stress}}
\end{align}
where 
\begin{align}
{{\mu\quer}}{}_{(ij)k} :=
\mu_{(ij)k}-\frac{1}{3}\,\delta_{ij}\, \mu_{k}.
\end{align}
The antisymmetric part (9 components) $\mu_{[ij]k}$ can be called rotatory hyperstress,
its conventional name is moment stress (or couple stress).
Here $\mu_k:=\mu_{llk}$ is the dilatational hyperstress (3 components) and the symmetric 
traceless part (15 components) ${{\mu\quer}}{}_{(ij)k}$ denotes the shearing hyperstress (without moment).

In the case of material linearity, 
the strain energy density is given by 
\begin{align}
\label{w}
w&=\frac{1}{2}\,\gamma_{ij}\, A_{ijkl}\gamma_{kl}
+\frac{1}{2}\,e_{ij}\, B_{ijkl} e_{kl}
+\frac{1}{2}\kappa_{ijk}\, C_{ijklmn}\kappa_{lmn}\nonumber\\ 
&\quad+\gamma_{ij}\, E_{ijkl} e_{kl}
+\gamma_{ij}\, F_{ijklm} \kappa_{klm}
+e_{ij}\, G_{ijklm}\kappa_{klm}
\end{align}
in terms of the geometric fields: 
the relative deformation (9 components) $\gamma_{ij}$, 
the micro-strain (6 components) $e_{ij}$
and the micro-wryness tensor (27 components) $\kappa_{ijk}$.

For a compatible material they read~\citep{Eringen99}
\begin{align}
&\gamma_{ij}=\pd_j u_i-\phi_{ij},\\
&e_{ij}=e_{ji}=\frac{1}{2}(\phi_{ij}+\phi_{ji}),\\
&\kappa_{ijk}=\pd_k\phi_{ij},
\end{align}
where $u_i$ is the displacement vector of the mass center of the particle,
called macro-displacement, 
and $\phi_{ij}$ denotes the micro-deformation tensor of the inner structure of
the particle.
It is obvious that the relative deformation is the difference between
the macro- and micro-deformation. 
For a compatible material, the macro-distortion is just the gradient 
of the macro-displacement: $\pd_j u_i$ .
The deformation of the directors is coded into
the micro-wryness tensor. 
Thus, $e_{ij}$ and $\kappa_{ijk}$ are determined by the 
micro-deformation, only.
For the relative distortion, $\gamma_{ij}$, 
we have the decomposition according to:
\begin{align}
\gamma_{ij}=\underbrace{\gamma_{[ij]}}_{\text{relative rotation}} 
+\underbrace{\gamma_{(ij)}}_{\text{relative strain}} ,
\end{align}
where the relative strain is the difference between the macro-strain,
$1/2(\pd_j u_i+\pd_i u_j)$,
 and the micro-strain $e_{ij}$.
Furthermore, $\gamma_{(ij)}$ can be split into
the dilatational part $\gamma:=\gamma_{ll}/3$ and a deviatoric part 
as follows
\begin{align}
\gamma_{(ij)}=
\delta_{ij}
\underbrace{\gamma}_{\text{relative dilatation}}
+\underbrace{
{{\gamma\quer}}{}_{ij}}_{\text{relative shear}}
\end{align}
where 
\begin{align}
{{\gamma\quer}}{}_{ij} :=\gamma_{(ij)}-
\delta_{ij}\, \gamma .
\end{align}
The micro-strain can be decomposed according to
\begin{align}
e_{(ij)}=
\delta_{ij}
\underbrace{e}_{\text{intrinsic dilatation}}
+\underbrace{
{{e\quer}}{}_{ij}}_{\text{intrinsic shear}},
\end{align}
where $e:=e_{ll}/3$. 
The wryness tensor can be decomposed into
its antisymmetric part, its trace, and its symmetric traceless
part with respect to its first two indices $i$ and $j$
\begin{align}
\kappa_{ijk}=
\underbrace{\kappa_{[ij]k}}_{\text{intrinsic rotational wryness}}
+\ 
\delta_{ij}
\underbrace{ \kappa_{k}}_{\text{intrinsic dilatation wryness}}
+\underbrace{
{{\kappa\quer}}{}_{(ij)k}}_{\text{intrinsic shear wryness}}
\end{align}
with $\kappa_k:=\kappa_{llk}/3$.

In Eq.~(\ref{w}) the tensors $A_{ijkl}$, $B_{ijkl}$, $C_{ijklmn}$, $E_{ijkl}$, $F_{ijklm}$
and $G_{ijklm}$ are constitutive tensors.
In general,
the constitutive tensors fulfill the symmetry relations:
\begin{align}
\label{sym}
&A_{ijkl}=A_{klij},\quad
B_{ijkl}=B_{klij}=B_{jikl}=B_{ijlk},\quad
C_{ijklmn}=C_{lmnijk},\nonumber\\
&E_{ijkl}=E_{jikl},\quad
G_{ijklm}=G_{jiklm}.
\end{align}
Only for a compatible material,
one has the additional symmetry
\begin{align}
\label{symm-ad}
C_{ijklmn}=C_{ijnlmk}.
\end{align}
which is used by \citet{Chen}.
If $\kappa_{ijk}$ is not a gradient, Eq.~(\ref{symm-ad}) is 
not valid.

The energy within a fixed volume of the body, over which the integrals will be taken
is 
\begin{align}
\label{W}
W=\int w\, \dd^3 x .
\end{align}
The constitutive equations where the material tensors vary with 
position read (see also~\citet{Eringen99} for homogeneous media): 
\begin{align}
\label{CE1}
\sigma_{ij}&=\frac{\pd w}{\pd \gamma_{ij}}
=A_{ijkl}\gamma_{kl}+E_{ijkl} e_{kl}+F_{ijklm} \kappa_{klm} ,\\
\label{CE2}
s_{ij}&=\frac{\pd w}{\pd e_{ij}}
=E_{klij}\gamma_{kl}+B_{ijkl} e_{kl}+G_{ijklm} \kappa_{klm} ,\\
\label{CE3}
\mu_{ijk}&= \frac{\pd w}{\pd \kappa_{ijk}}
=F_{lmijk}\gamma_{lm}+G_{lmijk} e_{lm}
+C_{ijklmn}\kappa_{lmn}
\end{align}
with $s_{ij}=s_{ji}$.

\subsection{Incompatibility conditions and Bianchi identities in 
micromorphic elasticity}
If the elastic tensors $\gamma_{ij}$, $e_{ij}$ and $\kappa_{ijk}$ are 
incompatible, which means they are not gradients, 
the total ones are decomposed independently into elastic and plastic
contributions (see, e.g., \citet{FS}):
\begin{align}
\label{deco1}
&\gamma^{\text{T}}_{ij}:=\pd_j u_i-\phi_{ij}=\gamma_{ij}+\gamma^{\text{P}}_{ij},\\
\label{deco2}
&e_{ij}^{\text{T}}:=\frac{1}{2}(\phi_{ij}+\phi_{ji})=e_{ij}+e_{ij}^{\text{P}},\\
\label{deco3}
&\kappa^{\text{T}}_{ijk}:=\pd_k\phi_{ij}=\kappa_{ijk}+\kappa_{ijk}^{\text{P}}.
\end{align}
Also the plastic parts $\gamma_{ij}^{\text{P}}$, $e_{ij}^{\text{P}}$ and $\kappa_{ij}^{\text{P}}$ are 
incompatible.
The total fields satisfy the compatibility conditions~\citep{Eringen99}:
\begin{align}
\label{CC1}
\epsilon_{jkl}(\pd_k\gamma^{\text{T}}_{il}+\kappa^{\text{T}}_{ilk})&=0,\\
\label{CC2}
\epsilon_{jkl}\pd_k\kappa^{\text{T}}_{inl}&=0,\\
\label{CC3}
-2\pd_k e^{\text{T}}_{ij}+\kappa^{\text{T}}_{ijk}+\kappa^{\text{T}}_{jik}&=0.
\end{align} 
Eventually, the elastic and plastic fields have to fulfill 
incompatibility conditions.
In linear micromorphic elasticity the incompatibility equations 
for the elastic fields are
given by
\begin{align}
\label{IC1}
\alpha_{ij}&=\epsilon_{jkl}(\pd_k\gamma_{il}+\kappa_{ilk}),\\
\label{IC2}
\Theta_{inj}&=\epsilon_{jkl}\pd_k\kappa_{inl},\\
\label{IC3}
Q_{ijk}&=-2\pd_k e_{ij}+\kappa_{ijk}+\kappa_{jik},
\end{align}
and for the plastic fields
\begin{align}
\label{IC1-p}
\alpha_{ij}&=-\epsilon_{jkl}(\pd_k\gamma^{\text{P}}_{il}+\kappa^{\text{P}}_{ilk}),\\
\label{IC2-p}
\Theta_{inj}&=-\epsilon_{jkl}\pd_k\kappa^{\text{P}}_{inl},\\
\label{IC3-p}
Q_{ijk}&=2\pd_k e^{\text{P}}_{ij}-\kappa^{\text{P}}_{ijk}-\kappa^{\text{P}}_{jik}.
\end{align} 
First of all, we note that the tensors $\Theta_{ijk}$ and $Q_{ijk}$ 
are determined by the deformation of the micro-structure. 
Only, $\alpha_{ij}$ contains the information of
the macro-deformation coded into the relative deformation $\gamma_{ij}$.
$\alpha_{ij}$ is the
dislocation density tensor which is identified with the 
linearized version of the Cartan torsion tensor
$T_{ijk}$ according to
\begin{align}
\alpha_{ij}=\frac{1}{2}\, \epsilon_{jkl} T_{ikl}.
\end{align} 
It has 9 independent components.

$\Theta_{ijk}$ 
is a tensor of third rank obtained from the (linearized) Riemann-Cartan curvature 
$R_{ijkl}$, which is a fourth rank curvature tensor, as follows
\begin{align}
\Theta_{ijk}=\frac{1}{2}\, \epsilon_{kmn} R_{ijmn}.
\end{align}
It possesses 27 independent components.
The curvature tensor $\Theta_{ijk}$ can be decomposed 
into its symmetric and skewsymmetric parts 
according to
\begin{align}
\Theta_{ijk}=\Theta_{(ij)k}+\Theta_{[ij]k}=
\Theta_{(ij)k}-\epsilon_{ijl}\Theta_{lk},
\end{align}
where (the minus sign is convention) 
\begin{align}
\Theta_{ij}=-\frac{1}{2}\,\epsilon_{ikl}\Theta_{klj}=
-\frac{1}{4}\, \epsilon_{ikl}\epsilon_{jmn}\, R_{klmn} 
\end{align}
is the part which is called disclination density tensor.
It is the three-dimensional (negative) Einstein tensor of $R_{klmn}$.
$\Theta_{[ij]k}$ is a ``rotational'' curvature and 
possesses 9 components. 
Whereas $\Theta_{(ij)k}$ is a ``strain'' curvature and
has 18 independent components.
In addition, the curvature $\Theta_{(ij)k}$ may be decomposed
\begin{align}
\label{curv2}
\Theta_{(ij)k}={{\Theta\quer}}{}_{(ij)k}
+\delta_{ij}\, \Theta_k .
\end{align}
The first part in Eq.~(\ref{curv2}) is a curvature of shear type, which has
15 components, and the trace $\Theta_k:=\Theta_{llk}/3$ is the dilatational
part with 3 components of the curvature. The dilatation part
$\Theta_k$ is called by \citet{Weyl} ``distance curvature''.

$Q_{ijk}$ is the tensor 
of nonmetricity ($Q_{ijk}=Q_{jik}$). Therefore, the micro-strain $e_{ij}$ is nonmetric.
One may interprete $Q_{ijk}$ as a measure of (intrinsic) point defects
disturbing the metricity~\citep{Kroener81,Kroener92}. 
Such point defects can be vacancies, self-intersticials and shear defects.
The nonmetricity tensor gives rise to 
a curvature tensor that has a symmetric part $\Theta_{(ij)k}$.
The nonmetricity tensor possesses 18 components as $\Theta_{(ij)k}$.
The nonmetricity can be decomposed into a traceless part and its trace
according to
\begin{align}
\label{Q2}
Q_{ijk}={{Q\quer}}{}_{ijk}
+\delta_{ij}\, Q_k ,
\end{align}
where $Q_k:=Q_{llk}/3$ is the so-called Weyl covector. 
$Q_k$ is related to (local) scale transformations, whereas ${Q\quer}_{ijk}$ 
is related to shear transformations, accordingly, it is a generic field
obstructing local rotation invariance.
Thus, the first 
part of Eq.~(\ref{Q2}) corresponds to defects of shear type and 
the Weyl covector describes defects of dilatation type.

From the differential-geometrical point of view $\gamma_{ij}$, $\kappa_{ijk}$ and
$e_{ij}$ are a coframe (9 components), connection (27 components) 
and a metric (6 components), respectively. These 
fields are independent field variables in a microcontinuum field theory.

The torsion, curvature and the nonmetricity fulfill the following 
Bianchi identities\footnote{Sometimes the third Bianchi identity is 
called the zeroth Bianchi identity, e.g.,~\citet{Hehl95,hehl97}.}
 \citep{schouten,Pov}
\begin{alignat}{2}
\pd_j\alpha_{ij}&=\Theta_{ijj}
&&\qquad\text{1st identity},\\
\pd_j\Theta_{inj}&=0
&&\qquad\text{2nd identity},\\
\epsilon_{klm}\pd_l Q_{ijm}&=2\Theta_{(ij)k}
&&\qquad\text{3rd identity}
\end{alignat}
which must be satisfied by $\alpha_{ij}$, $\Theta_{ijk}$ and $Q_{ijk}$.
The first Bianchi identity means that dislocations can interact with 
disclinations and point defects. The curvature $\Theta_{ijk}$ is 
divergenceless in the last index. 
From the third Bianchi identity it can be seen that the part of the curvature
$\Theta_{(ij)k}$ is given in terms of the nonmetricity tensor which
describes point defects.
Thus, $\Theta_{(ij)k}$ only emerges if nonmetricity is admitted.

If dislocations are only present in a micromorphic medium,
the nonmetricity tensor and the curvature tensor are zero and 
the dislocation density tensor reduces to
\begin{align}
\alpha_{ij}=\epsilon_{jkl}\pd_k \beta_{il}=-\epsilon_{jkl}\pd_k \beta_{il}^{\text{P}},
\end{align}
where $\beta^{\text{T}}_{ij}:=\pd_j u_i=\beta_{ij}+\beta^{\text{P}}_{ij}$
is the macro-displacement gradient.
Thus, it fulfills the Bianchi identity
\begin{align}
\pd_j\alpha_{ij}=0.
\end{align}
In such a dislocation theory of a micromorphic medium
the dislocation density tensor is completely determined by the 
macro-deformation like in classical dislocation theory~\citep{Kroener}.

\subsection{The Eshelby stress tensor for micromorphic elasticity}
In this subsection, we want to derive the Eshelby stress tensor and the configurational
forces for an anisotropic, inhomogeneous and incompatible micromorphic material
with external forces and couples. First we calculate the Eshelby stress tensor.
Its divergence gives the configurational forces. Such configurational
forces are of importance for the interaction between defects like dislocations
and disclinations.

First, we consider the simplest case -- a compatible and homogeneous micromorphic 
elasticity. 
In this case, one has~\citep{LA06}
\begin{align}
P_{ij}=w\delta_{ij}-\frac{\pd w}{\pd (\pd_i \BPhi)}(\pd_j \BPhi),\qquad
\pd_i P_{ij}=0
\end{align}
with the fields
\begin{align}
\BPhi=(u_i,\phi_{ij}).
\end{align}
So, we obtain~\citep{LA06}
\begin{align}
\label{P-2}
P_{ij}&=w\delta_{ij} -\frac{\pd w}{\pd (\pd_i u_l)}(\pd_j u_l)
-\frac{\pd w}{\pd (\pd_i \phi_{kl})}(\pd_j \phi_{kl})\nonumber\\
&=w\delta_{ij} -\sigma_{li}\pd_{j}u_l-\mu_{kli}\pd_{j}\phi_{kl} .
\end{align}

Of course, the following questions arise:
What are the configurational forces in micro\-morphic elasticity and 
what is the generalization of  Eq.~(\ref{P-2}) for an incompatible and 
non-homogeneous micromorphic elasticity?

Following the procedure of~\citet{Kirchner99}, we construct the Eshelby stress 
(or static energy-momentum) tensor for an incompatible and non-homogeneous 
micromorphic medium. 
From Eqs.~(\ref{w}) and (\ref{W}) we obtain for the infinitesimal 
variation $\delta$
\begin{align}
\label{W-var1}
\delta W&=\int\Big\{ 
\frac{1}{2} \gamma_{ij}[\delta A_{ijkl}]\gamma_{kl}
+ A_{ijkl} \gamma_{kl} [\delta \gamma_{ij}]
+\frac{1}{2}e_{ij}[\delta B_{ijkl}] e_{kl}
+ B_{ijkl} e_{kl} [\delta e_{ij}]\nonumber\\
&\qquad\quad
+\frac{1}{2}\kappa_{ijk}[\delta C_{ijklmn}]\kappa_{lmn}
+ C_{ijklmn} \kappa_{lmn} [\delta \kappa_{ijk}]
\nonumber\\
&\qquad\quad
+\gamma_{ij}[\delta E_{ijkl}] e_{kl}
+ E_{ijkl} e_{kl} [\delta \gamma_{ij}]
+ E_{ijkl} \gamma_{ij} [\delta e_{kl}]
\nonumber\\
&\qquad\quad
+\gamma_{ij}[\delta F_{ijklm}] \kappa_{klm}
+ F_{ijklm} \kappa_{klm} [\delta \gamma_{ij}]
+ F_{ijklm} \gamma_{ij} [\delta \kappa_{klm}]
\nonumber\\
&\qquad\quad
+e_{ij}[\delta G_{ijklm}] \kappa_{klm}
+ G_{ijklm} \kappa_{klm} [\delta e_{ij}]
+ G_{ijklm} e_{ij} [\delta \kappa_{klm}]
\Big\}\, \dd^3 x  .
\end{align}
With Eqs.~(\ref{CE1})--(\ref{CE3}) there remains
\begin{align}
\delta W&=\int\Big\{ 
\sigma_{ij}[\delta \gamma_{ij}]+s_{ij}[\delta e_{ij}]+
\mu_{ijk}[\delta \kappa_{ijk}]+
\frac{1}{2}\gamma_{ij}[\delta A_{ijkl}]\gamma_{kl}
+\frac{1}{2}e_{ij}[\delta B_{ijkl}] e_{kl}\\
&\qquad\quad
+\frac{1}{2}\kappa_{ijk}[\delta C_{ijklmn}]\kappa_{lmn}
+\gamma_{ij}[\delta E_{ijkl}] e_{kl}
+\gamma_{ij}[\delta F_{ijklm}] \kappa_{klm}
+e_{ij}[\delta G_{ijklm}] \kappa_{klm}
\Big\}\, \dd^3 x  \nonumber.
\end{align}
Now, we specify the variational operator to be translational:
\begin{align}
\delta=(\delta x_k) \pd_k .
\end{align}
So, it gives the expression
\begin{align}
\label{W3}
\delta W&=\int\Big\{ 
\sigma_{ij}[\pd_k \gamma_{ij}]+s_{ij}[\pd_k e_{ij}]+
\mu_{ijk}[\pd_k \kappa_{ijk}]
+\frac{1}{2}\gamma_{ij}[\pd_k A_{ijlm}]\gamma_{lm}
+\frac{1}{2}e_{ij}[\pd_k B_{ijlm}] e_{lm}\nonumber\\
&\qquad\quad
+\frac{1}{2}\kappa_{ijl}[\pd_k C_{ijlmnp}]\kappa_{mnp}
+\gamma_{ij}[\pd_k E_{ijlm}] e_{lm}
+\gamma_{ij}[\pd_k F_{ijlmn}] \kappa_{lmn}\nonumber\\
&\qquad\qquad
+e_{ij}[\pd_k G_{ijlmn}] \kappa_{lmn}
\Big\}(\delta x_k)\, \dd^3 x  .
\end{align}  
Using the equilibrium equations (\ref{EC1}) and (\ref{EC2}), and the incompatibility conditions
(\ref{IC1})--(\ref{IC3}),
we rewrite Eq.~(\ref{W3}) in the following form
\begin{align}
\label{W-var-r}
\delta W&=\int\Big\{ 
\epsilon_{kjl} \sigma_{ij}\alpha_{il}+\sigma_{ij}\kappa_{ikj}
+\pd_j [\sigma_{ij}\gamma_{ik}]+f_i\gamma_{ik}
-\frac{1}{2}s_{ij}Q_{ijk}
+\epsilon_{kmn}\mu_{ijm}\Theta_{ijn}\nonumber\\
&\qquad\quad
+\pd_j[\mu_{imj}\kappa_{imk}]+l_{ij}\kappa_{ijk}
+\text{nonhomogeneous terms}
\Big\}(\delta x_k)\, \dd^3 x  .
\end{align}
On the left hand side of Eq.~(\ref{W-var1}) we write 
\begin{align}
\label{W-var-l}
\delta W=\int \delta w\, \dd^3 x=\int[\pd_k w](\delta x_k)\, \dd^3 x
=\int\pd_i[w \delta_{ik}](\delta x_k)\, \dd^3 x .
\end{align}     
One obtains 
\begin{align}
\label{W4}
&\int\Big\{ 
\epsilon_{kjl} \sigma_{ij}\alpha_{il}
+\epsilon_{kmn}\mu_{ijm}\Theta_{ijn}
-\frac{1}{2}s_{ij}Q_{ijk}
+f_i\gamma_{ik}
+l_{ij}\kappa_{ijk}
+\sigma_{ij}\kappa_{ikj}
\nonumber\\
&\qquad
+\frac{1}{2}\gamma_{ij}[\pd_k A_{ijlm}]\gamma_{lm}
+\frac{1}{2}e_{ij}[\pd_k B_{ijlm}] e_{lm}
+\frac{1}{2}\kappa_{ijl}[\pd_k C_{ijlmnp}]\kappa_{mnp}
\nonumber\\
&\qquad
+\gamma_{ij}[\pd_k E_{ijlm}] e_{lm}
+\gamma_{ij}[\pd_k F_{ijlmn}] \kappa_{lmn}
+e_{ij}[\pd_k G_{ijlmn}] \kappa_{lmn}
\Big\}\, \dd^3 x
\nonumber\\  
&\hspace{6cm}
=\int\pd_i[w\delta_{ik}-\sigma_{li}\bar{\gamma}_{lk}-\mu_{jli}\kappa_{jlk}]\, \dd^3 x=J_k ,
\end{align}
where $\bar\gamma_{ij}=\pd_j u_i-\gamma^{\text{P}}_{ij}$.
The first part contains the configurational force densities.
The integrand of the second integral in Eq.~(\ref{W4}) is the divergence
of the Eshelby stress tensor (or static energy-momentum tensor)
\begin{align}
\label{P}
P_{ij}=w\delta_{ij}-\sigma_{li}\bar{\gamma}_{lj}-\mu_{kli}\kappa_{klj}.
\end{align}
Now, more important, its divergence gives the configurational 
(material), thermodynamic forces acting on the sources and inhomogeneities.
The result is 
\begin{align}
F_j=\pd_i P_{ij},
\end{align}
with
\begin{align}
\label{F}
F_j &=\epsilon_{jkl}\sigma_{ik}\alpha_{il}+\epsilon_{jkl}\mu_{mnk}\Theta_{mnl}
-\frac{1}{2} s_{mn}Q_{mnj}-\sigma_{kl}\kappa^{\text{P}}_{kjl}
+f_i\bar\gamma_{ij}+l_{kl}\kappa_{klj}+f^{\text{inh}}_j ,
\end{align}
where the inhomogeneities force density is due to the gradient of the
elastic tensors (see also~\citep{Eshelby51,Maugin93}):
\begin{align}
\label{f-inh1}
f^{\text{inh}}_j&=
\frac{1}{2}\gamma_{ik}[\pd_j A_{iklm}]\gamma_{lm}
+\frac{1}{2}e_{ik}[\pd_j B_{iklm}] e_{lm}
+\frac{1}{2}\kappa_{ikm}[\pd_j C_{iklmnp}]\kappa_{mnp}\nonumber\\
&\quad
+\gamma_{ik}[\pd_j E_{iklm}] e_{lm}
+\gamma_{ik}[\pd_j F_{iklmn}] \kappa_{lmn}
+e_{ik}[\pd_j G_{iklmn}] \kappa_{lmn} .
\end{align}
Equation~(\ref{F}) is a sum of configurational force densities:
the generalized Peach-Koehler force density on a dislocation density $\alpha_{il}$ in the presence
of the force stress $\sigma_{ik}$, 
the generalized Mathisson-Papapetrou force density on a 
defect density $\Theta_{mnl}$ in the presence of the 
hyper stress $\mu_{mnk}$~(see also~\citep{Maugin93}).
This part is 
known from gauge theory (see, e.g., \citep{Hehl95,Gairola81}).
A contribution from the nonmetricity is a new feature in micromorphic
elasticity; it arises because of the incompatible micro-strain and
micro-wryness.
Furthermore, we have
a generalized Cherepanov force density on a body force $f_i$ in the presence 
of the distortion $\bar\gamma_{ij}$, 
the force density on a body couple tensor $l_{kl}$ in presence 
of the elastic wryness $\kappa_{ij}$, 
the force density on the force stress $\sigma_{kl}$ in
presence of $\kappa^{\text{P}}_{kjl}$,
and the force densities on inhomogeneities 
in micromorphic elasticity $f_j^{\text{inh}}$
like the force density on an elastic inhomogeneity
derived by~\citet{Eshelby51}.   
Without sources and for a compatible and homogeneous medium, 
the Eshelby stress tensor (\ref{P})
is in agreement with Eq.~(\ref{P-2}). 

\section{Theory of microstretch elasticity}
\setcounter{equation}{0}
\subsection{The field equations}
In microstretch elasticity~\citep{Eringen92,Eringen99} we only have dilatational degrees of freedom 
in addition to the rotational and translational ones. 
Sometimes such a theory with independent rotational and dilatational fields
is called micropolar-dilatation elasticity~\citep{Markov}. 
It is a generalization of micropolar elasticity and a special case
of the micromorphic elasticity. Thus, all equations for microstretch elasticity 
can be reduced from the micromorphic ones by setting the shear parts of
the micro-fields to be zero. In microstretch elasticity, one has directors 
with stretch and rotation only and no micro-shears of them.

If we use the decomposition  
\begin{align}
\mu_{ijk}&=\frac{1}{3}\,\delta_{ij}\mu_k-\frac{1}{2}\, \epsilon_{ijl}\mu_{lk},
\quad\mu_{ij}=-\epsilon_{ilk}\mu_{klj},\\
l_{ij}&=\frac{1}{3}\,\delta_{ij} l-\frac{1}{2}\, \epsilon_{ijk} l_{k},
\quad l_{i}=-\epsilon_{ijk} l_{jk},
\end{align}
then we get from micromorphic equilibrium equations~(\ref{EC1}) and (\ref{EC2})
the equilibrium conditions in microstretch elasticity 
\begin{align}
\label{EC1-ms}
&\pd_j\sigma_{ij}+f_i=0 ,\\
\label{EC2-ms}
&\pd_j\mu_{ij}-\epsilon_{ijk}\sigma_{jk}+l_{i}=0,\\
\label{EC3-ms}
&\pd_k\mu_{k}+\sigma_{}-s_{}+l_{}=0.
\end{align}
Here $\mu_k$ is the intrinsic dilatational stress or microstretch vector
and $\mu_{ij}$ is the moment stress tensor,
$l$ is the body microstretch force and $l_i$ the body couple.
Eqs.~(\ref{EC1-ms}) and (\ref{EC2-ms}) have the same form as the
balance equations in micropolar elasticity. Eq.~(\ref{EC3-ms}) is the 
additional microstretch equilibrium condition which 
describes the dilatational balance condition.

Using the passage to microstretch elasticity in Eq.~(\ref{w})
with
\begin{align}
\kappa_{ijk}&=\delta_{ij}\kappa_k-\epsilon_{ijl}\kappa_{lk},
\quad\kappa_{ij}=-\frac{1}{2}\,\epsilon_{ilk}\kappa_{klj},\\
e_{ij}&=\delta_{ij} e,
\end{align}
the strain energy is given by
\begin{align}
\label{w-ms}
w&=\frac{1}{2}\,\gamma_{ij}\, A_{ijkl}\gamma_{kl}
+\frac{1}{2}\,e_{}\, B_{} e_{}
+\frac{1}{2}\kappa_{ij}\, C_{ijkl}\kappa_{kl}
+\frac{1}{2}\kappa_{i}\, C_{ij}\kappa_{j}
+\frac{1}{2}\,\kappa_{i}\, C_{ijk}\kappa_{jk}
+\frac{1}{2}\kappa_{ij}\, C_{ijk}\kappa_{k}
\nonumber\\ 
&\quad+\gamma_{ij}\, E_{ij} e_{}
+\gamma_{ij}\, F_{ijkl} \kappa_{kl}
+\gamma_{ij}\, F_{ijk} \kappa_{k}
+e_{}\, G_{ij}\kappa_{ij}
+e_{}\, G_{i}\kappa_{i},
\end{align}
with
\begin{align}
\label{sym-ms}
&B=B_{kkll},\quad
C_{ij}=C_{kkillj},\quad
C_{ijk}=-\epsilon_{imn} C_{mnjllk},\quad
C_{ijkl}=\epsilon_{imn} C_{mnjpql}\epsilon_{pqk},\\
&E_{ij}=E_{ijkk},\quad
F_{ijk}=F_{ijllk},\quad
F_{ijkl}=-F_{ijmnl}\epsilon_{mnk},\quad
G_{i}=G_{kklli},\quad
G_{ij}=-G_{kkmnj}\epsilon_{imn}.\nonumber
\end{align}
Thus, the microstretch constitutive moduli are 
reduced from the micromorphic constitutive moduli.
From Eqs.~(\ref{sym}) and (\ref{sym-ms}) 
the following symmetry relations are obtained
\begin{align}
\label{sym-ms2}
A_{ijkl}=A_{klij},\quad
C_{ijkl}=C_{klij},\quad
C_{ij}=C_{ji},\quad
C_{ijk}=C_{kij} .
\end{align}
Using Eq.~(\ref{w-ms}), the linear constitutive relations are explicitly
\begin{align}
\label{CE1-ms}
\sigma_{ij}&=\frac{\pd w}{\pd \gamma_{ij}}
=A_{ijkl}\gamma_{kl}+E_{ij} e_{}+F_{ijkl} \kappa_{kl}
+F_{ijk} \kappa_{k} ,\\
\label{CE2-ms}
s-\sigma&=\frac{\pd w}{\pd \phi}
=(E_{kl}-A_{kl})\gamma_{kl}+(B-E) e+(G_{kl}-F_{kl}) \kappa_{kl}
+(G_l-F_l)\kappa_l ,\\
\label{CE3-ms}
\mu_{ij}&= \frac{\pd w}{\pd \kappa_{ij}}
=F_{klij}\gamma_{kl}+G_{ij} e_{}+C_{ijkl}\kappa_{kl}+C_{kij}\kappa_{k},\\
\label{CE-ms}
\mu_{i}&= \frac{\pd w}{\pd \kappa_{i}}
=F_{kli}\gamma_{kl}+G_{i} e_{}+C_{ikl}\kappa_{kl}+C_{ki}\kappa_{k},
\end{align}
with $A_{ij}=A_{ijkk}=A_{kkij}$.

\subsection{Incompatibility conditions and Bianchi identities 
in micro\-stretch elasticity}
The total fields are again considered to be a sum of elastic and plastic 
incompatible parts. 
Without micro-shears we obtain from Eqs.~(\ref{deco1})--(\ref{deco3})
the following decomposition into elastic and plastic parts:
\begin{align}
&\gamma^{\text{T}}_{ij}:=\pd_j u_i+\epsilon_{ijk}\phi_k-\delta_{ij}\phi=\gamma_{ij}+\gamma^{\text{P}}_{ij},\\
&e^{\text{T}}:=\phi_{}=e+e^{\text{P}},\\
&\kappa^{\text{T}}_{ij}:=\pd_j\phi_{i}=\kappa_{ij}+\kappa_{ij}^{\text{P}},\\
&\kappa^{\text{T}}_{j}:=\pd_j\phi=\kappa_{j}+\kappa_{j}^{\text{P}}.
\end{align}
The total fields must satisfy the following compatibility conditions:
\begin{align}
\label{CC1-ms}
\epsilon_{jkl}(\pd_k\gamma^{\text{T}}_{il}+\epsilon_{ikm}\kappa^{\text{T}}_{ml}-\delta_{ik}\kappa^{\text{T}}_l)&=0,\\
\label{CC2a-ms}
\epsilon_{jkl}\pd_k\kappa^{\text{T}}_{il}&=0,\\
\label{CC2b-ms}
\epsilon_{jkl}\pd_k\kappa^{\text{T}}_{l}&=0,\\
\label{CC3-ms}
-\pd_k e^{\text{T}}+\kappa^{\text{T}}_{k}&=0.
\end{align} 
It is important to note that these compatibility equations do not
exactly agree with the ones given by~\citet{Eringen99}.
He claimed that the compatibility conditions of microstretch elasticity 
are identical to those of micropolar elasticity.
In fact, Eqs.~(\ref{CC2b-ms}) and (\ref{CC3-ms}) and the last part in (\ref{CC1-ms})
are missing in his conditions. 
But, on the other hand, in microstretch elasticity we have four
fields, two fields more than in micropolar elasticity. 
For each one we have a compatibility condition. Thus, we need four
compatibility conditions in microstretch elasticity. 

The generalization of the compatibility conditions for the total fields
are the incompatibility
conditions for the (incompatible) elastic and plastic fields.
Eventually, the incompatibility conditions read for elastic fields
\begin{align}
\label{IC1-ms}
\alpha_{ij}&=\epsilon_{jkl}(\pd_k\gamma_{il}+\epsilon_{ikm}\kappa_{ml}-\delta_{ik}\kappa_l),\\
\label{IC2a-ms}
\Theta_{ij}&=\epsilon_{jkl}\pd_k\kappa_{il},\\
\label{IC2b-ms}
\Theta_{j}&=\epsilon_{jkl}\pd_k\kappa_{l},\\
\label{IC3-ms}
Q_k&=-2(\pd_k e-\kappa_{k}),
\end{align} 
and for the plastic fields 
\begin{align}
\label{IC1-p-ms}
\alpha_{ij}&=-\epsilon_{jkl}(\pd_k\gamma^{\text{P}}_{il}+\epsilon_{ikm}\kappa^{\text{P}}_{ml}-\delta_{ik}\kappa^{\text{P}}_l),\\
\label{IC2a-p-ms}
\Theta_{ij}&=-\epsilon_{jkl}\pd_k\kappa^{\text{P}}_{il},\\
\label{IC2b-p-ms}
\Theta_{j}&=-\epsilon_{jkl}\pd_k\kappa^{\text{P}}_{l},\\
\label{IC3-p-ms}
Q_k&=2(\pd_k e^{\text{P}}-\kappa^{\text{P}}_{k}).
\end{align}

By differentiating Eqs.~(\ref{IC1-ms})--(\ref{IC3-ms}) and 
Eqs.~(\ref{IC1-p-ms})--(\ref{IC3-p-ms}) we obtain
the Bianchi identities of microstretch elasticity
\begin{align}
\pd_j\alpha_{ij}&=\epsilon_{imk}\Theta_{mk}+\Theta_i
,\\
\pd_j\Theta_{ij}&=0 , 
\\
\pd_j\Theta_{j}&=0, \\
\epsilon_{klm}\pd_l Q_{m}&=2\Theta_{k},
\end{align}
which must be fulfilled by $\alpha_{ij}$, $\Theta_{ij}$, $\Theta_{j}$ 
and $Q_{j}$.
These Bianchi identities are in agreement with the ones 
given by~\citet{Pov}.

\subsection{The Eshelby stress tensor for micro\-stretch elasticity}
In microstretch elasticity, we get from Eq.~(\ref{P}) the following
Eshelby stress tensor
\begin{align}
\label{P-ms}
P_{ij}=w\delta_{ij}-\sigma_{li}\bar\gamma_{lj}-\mu_{li}\kappa_{lj}-\mu_i\kappa_j,
\end{align}
and from Eq.~(\ref{F}) we obtain the following expression for 
the configurational forces
\begin{align}
\label{F-ms}
F_j &=\epsilon_{jkl}\sigma_{ik}\alpha_{il}
+\epsilon_{jkl}\mu_{ik}\Theta_{il}
-\epsilon_{jki}\sigma_{kl}\kappa^{\text{P}}_{il}+f_i\bar\gamma_{ij}+l_{i}\kappa_{ij}
+\epsilon_{jkl}\mu_{k}\Theta_{l}-\frac{1}{2} s_{}Q_{j}
-\sigma_{jl}\kappa^{\text{P}}_{l}+l_{}\kappa_{j}+f^{\text{inh}}_j,
\end{align}
with
\begin{align}
\label{f-inh2}
f^{\text{inh}}_j&=
\frac{1}{2}\gamma_{ik}[\pd_j A_{iklm}]\gamma_{lm}
+\frac{1}{2}e_{}[\pd_j B_{}] e_{}
+\frac{1}{2}\kappa_{ik}[\pd_j C_{iklm}]\kappa_{lm}
+\frac{1}{2}\kappa_{i}[\pd_j C_{ik}]\kappa_{k}
\nonumber\\
&\quad
+\frac{1}{2}\kappa_{i}[\pd_j C_{ikl}]\kappa_{kl}
+\frac{1}{2}\kappa_{ik}[\pd_j C_{ikl}]\kappa_{l}
+\gamma_{ik}[\pd_j E_{ik}] e_{}
+\gamma_{ik}[\pd_j F_{iklm}] \kappa_{lm}
+\gamma_{ik}[\pd_j F_{ikl}] \kappa_{l}\nonumber \\
&\quad
+e_{}[\pd_j G_{ik}] \kappa_{ik}+e_{}[\pd_j G_{i}] \kappa_{i} ,
\end{align}
which consists of 20 parts: 9 source terms and incompatibilities
(dislocations, disclinations, point defects) 
and 11 nonhomogeneous terms.
Without source and nonhomogeneous terms and for the 
compatible situation we obtain the divergenceless Eshelby stress tensor
\begin{align}
\label{P-ms-2}
P_{ij}=w\delta_{ij}-\sigma_{li}\pd_j u_l-\mu_{li}\pd_j \phi_l-\mu_i\pd_j\phi ,
\end{align}
which is in agreement with the expression found by~\citet{LA06}.

\section{Theory of micropolar elasticity}
\setcounter{equation}{0}
\subsection{The field equations}
In micropolar elasticity~\citep{Eringen90,Eringen99} we only have rotational
degrees of freedom in addition to the translational ones. 
Thus, it is a special case of microstretch elasticity.

Now, the equilibrium conditions read
\begin{align}
\label{EC1-me}
&\pd_j\sigma_{ij}+f_i=0 ,\\
\label{EC2-me}
&\pd_j\mu_{ij}-\epsilon_{ijk}\sigma_{jk}+l_{i}=0.
\end{align}
Here $\mu_{ij}$ is the moment stress tensor and $l_i$ the body couple.
The stored strain energy is given by
\begin{align}
\label{w-me}
w&=\frac{1}{2}\,\gamma_{ij}\, A_{ijkl}\gamma_{kl}
+\frac{1}{2}\kappa_{ij}\, C_{ijkl}\kappa_{kl}
+\gamma_{ij}\, F_{ijkl} \kappa_{kl}.
\end{align}
Thus, the microstretch constitutive moduli are 
reduced micromorphic constitutive moduli.
From Eqs.~(\ref{sym}) and (\ref{sym-ms}) 
and using Eq.~(\ref{w-ms}), the linear constitutive relations are explicitely
\begin{align}
\label{CE1-me}
\sigma_{ij}&=\frac{\pd w}{\pd \gamma_{ij}}
=A_{ijkl}\gamma_{kl}+F_{ijkl} \kappa_{kl},\\
\label{CE3-me}
\mu_{ij}&= \frac{\pd w}{\pd \kappa_{ij}}
=F_{klij}\gamma_{kl}+C_{ijkl}\kappa_{kl}.
\end{align}
Here the tensors $A_{ijkl}$ and $C_{ijkl}$ possess the symmetries~(\ref{sym-ms2}).

\subsection{Incompatibility conditions and Bianchi identities 
in micro\-polar elasticity}
The total fields are considered to be a sum of elastic and plastic 
incompatible parts, i.e.,
\begin{align}
&\gamma^{\text{T}}_{ij}:=\pd_j u_i+\epsilon_{ijk}\phi_k=\gamma_{ij}+\gamma^{\text{P}}_{ij},\\
&\kappa^{\text{T}}_{ij}:=\pd_j\phi_{i}=\kappa_{ij}+\kappa_{ij}^{\text{P}}.
\end{align}
The total fields must satisfy the following compatibility conditions:
\begin{align}
\label{CC1-me}
\epsilon_{jkl}(\pd_k\gamma^{\text{T}}_{il}+\epsilon_{ikm}\kappa^{\text{T}}_{ml})&=0,\\
\label{CC2a-me}
\epsilon_{jkl}\pd_k\kappa^{\text{T}}_{il}&=0.
\end{align} 
It is important to note that these compatibility equations do 
agree with the ones given by~\cite{Eringen99}.

The generalization of the compatibility conditions for the total fields
are the incompatibility
conditions for the (incompatible) elastic and plastic fields.
Eventually, the incompatibility conditions read for elastic 
fields~\citep{Schaefer,CE69}
\begin{align}
\label{IC1-me}
\alpha_{ij}&=\epsilon_{jkl}(\pd_k\gamma_{il}+\epsilon_{ikm}\kappa_{ml}),\\
\label{IC2a-me}
\Theta_{ij}&=\epsilon_{jkl}\pd_k\kappa_{il},
\end{align} 
and for the plastic fields 
\begin{align}
\label{IC1-p-me}
\alpha_{ij}&=-\epsilon_{jkl}(\pd_k\gamma^{\text{P}}_{il}+\epsilon_{ikm}\kappa^{\text{P}}_{ml}),\\
\label{IC2a-p-me}
\Theta_{ij}&=-\epsilon_{jkl}\pd_k\kappa^{\text{P}}_{il}.
\end{align}

By differentiating Eqs.~(\ref{IC1-me})--(\ref{IC2a-p-me}) 
we obtain the Bianchi identities in micropolar elasticity~\citep{CE69}
\begin{align}
\pd_j\alpha_{ij}&=\epsilon_{imk}\Theta_{mk},\\
\pd_j\Theta_{ij}&=0 , 
\end{align}
which must be fulfilled by $\alpha_{ij}$ and $\Theta_{ij}$.
These Bianchi identities are in agreement with those given by~\citet{CE69}.

\subsection{The Eshelby stress tensor for micro\-polar elasticity}
In micropolar elasticity, we get from Eq.~(\ref{P}) the following
Eshelby stress tensor (see also~\citep{LK06}) 
\begin{align}
\label{P-me}
P_{ij}=w\delta_{ij}-\sigma_{li}\bar\gamma_{lj}-\mu_{li}\kappa_{lj},
\end{align}
and from Eq.~(\ref{F}) we obtain the following expression for 
the configurational forces
\begin{align}
\label{F-me}
F_j &=\epsilon_{jkl}\sigma_{ik}\alpha_{il}
+\epsilon_{jkl}\mu_{ik}\Theta_{il}
-\epsilon_{jki}\sigma^{\text{P}}_{kl}\kappa_{il}+f_i\bar\gamma_{ij}+l_{i}\kappa_{ij}
+f^{\text{inh}}_j,
\end{align}
where
\begin{align}
\label{f-inh3}
f^{\text{inh}}_j&=
\frac{1}{2}\gamma_{ik}[\pd_j A_{iklm}]\gamma_{lm}
+\frac{1}{2}\kappa_{ik}[\pd_j C_{iklm}]\kappa_{lm}
+\gamma_{ik}[\pd_j F_{iklm}] \kappa_{lm}
\end{align}
which consists of 8 parts: 5 source terms and incompatibilities 
(dislocations and disclinations) 
and 3 nonhomogeneous terms.
Without source and nonhomogeneous terms and for the 
compatible situation we obtain the divergenceless Eshelby stress tensor
as
\begin{align}
\label{P-me-2}
P_{ij}=w\delta_{ij}-\sigma_{li}\pd_j u_l-\mu_{li}\pd_j \phi_l .
\end{align}
The formula~(\ref{P-me-2}) is in agreement with the Eshelby stress tensor
given by~\citet{LM}.
The corresponding Eshelby stress tensor for finite theory of polar elasticity
has been given by~\citet{Maugin98}.

\section{Theory of dilatation elasticity}
\setcounter{equation}{0}
\subsection{The field equations}
In the dilatation elasticity~\citep{Markov74,Markov}, we have only the dilatation of the microstructure 
in addition to the macro displacement as independent degrees of freedom. 
Thus, the dilatation elasticity is obtained from the microstretch-elasticity
(or micropolar-dilatation elasticity) if we set the micro-rotational 
degrees of freedom to be zero.
Theories of such type were also introduced and considered by~\citet{NC79,CN83}.
These authors called it elasticity with voids. 
The equations of equilibrium are
\begin{align}
\label{EC1-dil}
&\pd_j\sigma_{ij}+f_i=0 ,\\
\label{EC3-dil}
&\pd_k\mu_{k}+\sigma_{}-s_{}+l_{}=0.
\end{align}
The strain energy is of the form
\begin{align}
\label{w-dil}
w&=\frac{1}{2}\,\gamma_{ij}\, A_{ijkl}\gamma_{kl}
+\frac{1}{2}\,e_{}\, B_{} e_{}
+\frac{1}{2}\kappa_{i}\, C_{ij}\kappa_{j}
+\gamma_{ij}\, E_{ij} e_{}
+\gamma_{ij}\, F_{ijk} \kappa_{k}
+e_{}\, G_{i}\kappa_{i} .
\end{align}
Then  we get for the stresses 
\begin{align}
\label{CE1-dil}
\sigma_{ij}&=\frac{\pd w}{\pd \gamma_{ij}}
=A_{ijkl}\gamma_{kl}+E_{ij} e_{}+F_{ijk} \kappa_{k} ,\\
\label{CE2-dil}
s-\sigma&=\frac{\pd w}{\pd \phi}
=(E_{kl}-A_{kl})\gamma_{kl}+(B-E) e+(G_l-F_l)\kappa_l ,\\
\label{CE-dil}
\mu_{i}&= \frac{\pd w}{\pd \kappa_{i}}
=F_{kli}\gamma_{kl}+G_{i} e_{}+C_{ki}\kappa_{k}.
\end{align}
The constitutive tensors have the same symmetries as in Eqs.~(\ref{sym-ms})
and (\ref{sym-ms2}).

\subsection{Incompatibility conditions and Bianchi identities 
in dilatation elasticity}
The total geometric fields are
\begin{align}
&\gamma^{\text{T}}_{ij}:=\pd_j u_i-\delta_{ij}\phi=\gamma_{ij}+\gamma^{\text{P}}_{ij},\\
&e^{\text{T}}:=\phi_{}=e+e^{\text{P}},\\
&\kappa^{\text{T}}_{j}:=\pd_j\phi=\kappa_{j}+\kappa_{j}^{\text{P}}.
\end{align}
Again, the total fields must satisfy the following compatibility conditions:
\begin{align}
\label{CC1-dil}
\epsilon_{jkl}(\pd_k\gamma^{\text{T}}_{il}-\delta_{ik}\kappa^{\text{T}}_l)&=0,\\
\label{CC2-dil}
\epsilon_{jkl}\pd_k\kappa^{\text{T}}_{l}&=0,\\
\label{CC3-dil}
-\pd_k e^{\text{T}}+\kappa^{\text{T}}_{k}&=0.
\end{align} 
The incompatibility conditions are given by
\begin{align}
\label{IC1-dil}
\alpha_{ij}&=\epsilon_{jkl}(\pd_k\gamma_{il}-\delta_{ik}\kappa_l),\\
\label{IC2-dil}
\Theta_{j}&=\epsilon_{jkl}\pd_k\kappa_{l},\\
\label{IC3-dil}
Q_k&=-2(\pd_k e-\kappa_{k}),
\end{align} 
and for the plastic fields 
\begin{align}
\label{IC1-p-dil}
\alpha_{ij}&=-\epsilon_{jkl}(\pd_k\gamma^{\text{P}}_{il}-\delta_{ik}\kappa^{\text{P}}_l),\\
\label{IC2b-p-dil}
\Theta_{j}&=-\epsilon_{jkl}\pd_k\kappa^{\text{P}}_{l},\\
\label{IC3-p-dil}
Q_k&=2(\pd_k e^{\text{P}}-\kappa^{\text{P}}_{k}).
\end{align}
Eventually, the Bianchi identities read
\begin{align}
\pd_j\alpha_{ij}&=\Theta_i, \\
\pd_j\Theta_{j}&=0, \\
\epsilon_{klm}\pd_l Q_{m}&=2\Theta_{k},
\end{align}
which must be fulfilled by $\alpha_{ij}$, $\Theta_{j}$ 
and $Q_{j}$.

\subsection{The Eshelby stress tensor for dilatation elasticity}
In dilatation elasticity, we get from Eq.~(\ref{P}) the following
Eshelby stress tensor
\begin{align}
\label{P-dil}
P_{ij}=w\delta_{ij}-\sigma_{li}\bar\gamma_{lj}-\mu_i\kappa_j,
\end{align}
and from Eq.~(\ref{F}) we obtain the following expression for 
the configurational forces
\begin{align}
\label{F-dil}
F_j &=\epsilon_{jkl}\sigma_{ik}\alpha_{il}
+f_i\bar\gamma_{ij}
+\epsilon_{jkl}\mu_{k}\Theta_{l}-\frac{1}{2} s_{}Q_{j}
-\sigma_{jl}\kappa^{\text{P}}_{l}+l_{}\kappa_{j}+f^{\text{inh}}_j,
\end{align}
with
\begin{align}
\label{f-inh4}
f^{\text{inh}}_j&=
\frac{1}{2}\gamma_{ik}[\pd_j A_{iklm}]\gamma_{lm}
+\frac{1}{2}e_{}[\pd_j B_{}] e_{}
+\frac{1}{2}\kappa_{i}[\pd_j C_{ik}]\kappa_{k}
+\gamma_{ik}[\pd_j E_{ik}] e_{}
+\gamma_{ik}[\pd_j F_{ikl}] \kappa_{l}
+e_{}[\pd_j G_{i}] \kappa_{i} ,
\end{align}
which consists of 12 parts: 6 source terms and incompatibilities 
(dislocations, point defects and/or microvoids)
and 6 nonhomogeneous terms.
Without source and nonhomogeneous terms and for the 
compatible situation we obtain the divergenceless Eshelby stress tensor
as
\begin{align}
\label{P-dul-2}
P_{ij}=w\delta_{ij}-\sigma_{li}\pd_j u_l-\mu_i\pd_j\phi .
\end{align}

If we set the dilatational degrees of freedom to be zero 
we would recover the equilibrium conditions, incompatibility equations,
Eshelby stress tensor and the configurational forces in
the anisotropic theory of nonhomogeneous elasticity given by~\citet{Kirchner99}.

\subsection*{Acknowledgement}
M.L. has been supported by an Emmy-Noether grant of the 
Deutsche Forschungsgemeinschaft (Grant No. La1974/1-2). 
G.A.M. benefits from a Max-Planck-Award for international cooperation 
(2002-2006).

\end{document}